 \documentclass[pmlr,twocolumn]{jmlr} 
 
 \DeclareUnicodeCharacter{0301}{\'{e}}

\usepackage{booktabs}
\usepackage[load-configurations=version-1]{siunitx} 

\usepackage{stfloats}
\usepackage{graphicx}
\usepackage{float}
\usepackage{tikz}
\usepackage{pgfplots}
\usepackage{makecell}
\usepackage{multirow}
\usepackage{warpcol}
\usepackage{threeparttable}
\usepackage[T1]{fontenc}

\theorembodyfont{\upshape}
\theoremheaderfont{\scshape}
\theorempostheader{:}
\theoremsep{\newline}

\jmlrvolume{LEAVE UNSET}
\jmlryear{2020}
\jmlrsubmitted{LEAVE UNSET}
\jmlrpublished{LEAVE UNSET}
\jmlrworkshop{Machine Learning for Health (ML4H) 2020} 

\title[Parkinsonian Chinese Speech Analysis]{Parkinsonian Chinese Speech Analysis towards Automatic Classification of Parkinson's Disease}

\author{
  \Name{Hao Fang} \Email{fangh18@mails.tsinghua.edu.cn}\\
  \Name{Chen Gong} \Email{gongc16@mails.tsinghua.edu.cn}\\
  \Name{Chen Zhang} \Email{zhangchen2020@tsinghua.edu.cn}\\
  \Name{Yanan Sui} \Email{ysui@tsinghua.edu.cn}\\
  \Name{Luming Li} \Email{lilm@tsinghua.edu.cn}\\
  \addr National Engineering Laboratory for Neuromodulation, Tsinghua University
}

\begin{document}

\maketitle

\begin{abstract}

Speech disorders often occur at the early stage of Parkinson's disease (PD). The speech impairments could be indicators of the disorder for early diagnosis, while motor symptoms are not obvious. In this study, we constructed a new speech corpus of Mandarin Chinese and addressed classification of patients with PD. We implemented classical machine learning methods with ranking algorithms for feature selection, convolutional and recurrent deep networks, and an end to end system. Our classification accuracy significantly surpassed state-of-the-art studies. The result suggests that free talk has stronger classification power than standard speech tasks, which could help the design of future speech tasks for efficient early diagnosis of the disease. Based on existing classification methods and our natural speech study, the automatic detection of PD from daily conversation could be accessible to the majority of the clinical population.

\end{abstract}


\section{Introduction}
\label{sec:intro}

Parkinson's disease (PD) is the second most common neurodegenerative disease in the world. The affected population keeps increasing as we expect an aging society. \citet{dorsey2007projected} estimated that by 2030, over eight million people will suffer from PD, while about half of them speaking Chinese. \citet{vaiciukynas2017parkinson} showed that intervention therapy in the early stage of PD could effectively alleviate the disease progression. Early diagnosis thus is crucial to lead to early medical treatments. However, it is still difficult to make early detection of patients with PD when their motor symptoms are not obvious. Speech disorders are common symptoms of PD. About 75\%-95\% of PD patients show speech impairments, such as mono-tone, mono-loudness, slurred speech, and loss of volume. These symptoms frequently occur at the onset of PD, long before the appearance of significant motor signs \citep{pawlukowska2018differences}. Therefore, speech deficits could be treated as potential indicators for early diagnosis of the majority of the clinical population \citep{rusz2011quantitative}. 

In this study, We aimed to assess the speech disorder and separate the clinical group from the healthy deploying various machine learning methods. We built a corpus of Chinese speech tasks, including one specific task of Chinese poem (structured sentences) reading. We implemented classical machine learning methods with ranking algorithms for feature selection, convolutional and recurrent deep networks, and an ene to end system. We investigated acoustic characteristics that could distinguish the speech disorders of patients with PD from healthy participants. The performances of our methods surpassed the state-of-the-art results.
Our results suggested that both classical machine learning methods with feature selection and advanced deep learning tools could effectively capture Parkinsonian speech characteristics. This work presented the first study on classifying PD patients from healthy subjects using Chinese speech signals to the best of our knowledge.

We further explored possible ways to optimize speech tasks towards better classification/diagnosis. Our free talk based image description task yielded better classification accuracy comparing to standard tasks, which is consistent with the physician's experience and has been proved in \citet{goberman2010characteristics}.
Our methods could be directly applied to clinical evaluations and potentially utilized for detecting patients with PD from daily speech. This assistive diagnostic system would be accessible to everyone when integrated into mobile applications.

\section{Related Work}
Previous studies showed the possibility of classification of patients with PD from their speech signals. The speech samples collected in these studies included sustained vowels, diadochokinetic (DDK), words, and sentences. Some studies achieved relatively good classification results from sustained vowels \citep{xu2018voiceprint,sakar2019comparative,gunduz2019deep}, mainly the vowel /a/. DDK, words, and sentences contained more varieties in pitch and rhythm compared to sustained vowels, comprehensively presenting more information about the speech disorders. Classifications of the PD group were reported from words and/or sentence tasks in different languages, including Hebrew \citep{hauptman2019identifying}, French \citep{jeancolas2019comparison}, Spanish \citep{lopez2019assessing}, German \citep{orozco2016automatic}, Czech \citep{orozco2016automatic}, and Lithuanian \citep{vaiciukynas2017parkinson}.

As features, time-frequency representations of acoustic signals are frequently used in machine learning studies for human speech. With the advantage of approximating the human auditory system's response \citep{logan2000mel}, Mel-Frequency Cepstral Coefficients (MFCCs) are one of the most commonly used features in these studies \citep{benba2015voiceprints,xu2018voiceprint,jeancolas2019comparison,2019Objective}.

Both classical machine learning and deep learning methods had been introduced in the classification of PD patients' speech. The Support Vector Machine (SVM) was widely selected and frequently performed well \citep{2018Multimodal,2018Unobtrusive}. Some studies also chose k nearest neighbor (KNN) and random forest (RF) to classify PD speech \citep{2013Collection,2017Can,polat2019hybrid}. In recent years, many studies chose deep learning methods, especially Convolutional Neural Network (CNN), as classifier \citep{J2017Convolutional,Juan2018A,gunduz2019deep}.

However, publicly available speech datasets and classification studies of patients with PD in Chinese are not found. Some people discussed the intonation contrast of PD patients speaking Mandarin Chinese \citep{liu2019prosodic} and Cantonese \citep{ma2010intonation}, without further attempts on classification.

\section{Data Collection}
\label{sec:data}

We collected a new Chinese speech corpus containing speech samples recorded from 34 patients with PD and 34 healthy controls (HCs) via different phones and DVs. This uncontrolled device and environment recording condition rather than well-controlled experimental settings was more applicable in daily life, making it accessible to the majority of the disorder group. It also helped to train classifiers which would be easier to generalize and more robust in real application. The set of PD patients included 20 males aged 45 to 73 years (mean $56.70\pm8.35$ years) and 14 females aged 41 to 68 years (mean $58.29\pm6.94$ years). All PD patients were clinically diagnosed by experienced neurologists. 

The HC group included 16 males aged 20-55 years (mean $41.22\pm14.89$ years) and 18 females aged 21-74 years (mean $48.55\pm11.65$ years). We noticed that our participant groups were not precisely matched in age and gender. However, \citet{sapir1999speech} proved that age and gender were not relevant to speech abnormalities. All subjects signed an informed consent form before their participation. 

Speech signals of each subject were sampled at 48kHz for three tasks: 

1) Image description (describe the image content after watching it for 30sec); 

2) DDK test (quickly repeat /lalala-tatata-dadada/ for three times);

3) Text reading (read two ancient Chinese poems, composed of eight seven-word sentences). 

\section{Methods}
\label{sec:methods}

\subsection{Preprocessing}
\label{sec:prepro}

We first removed non-speech episodes and non-subject speech episodes from all recordings. Then, speech segments were extracted for each task respectively, according to the following criteria: 

1) For image description, segments were extracted between speech pauses; 

2) For DDK test, each segment included one /lalala-tatata-dadada/ sample;

3) For text reading, each segment included one seven-word sentence. 

We excluded poor quality segments (e.g., containing significant noise, deviating from task requirements, etc.), and acquired 4820 speech segments in total for all 68 subjects (task 1: 2098, task 2: 773, task 3: 1949). Each segment was given a label indicating whether it belonged to PD patients or healthy subjects.

We calculated 128 MFCCs within 2000Hz frequency for each segment, using a sliding window of 2048 points (about 42.67msec) with an overlap of 75\%. Among all 128 MFCCs, the 5\textsuperscript{th} to 44\textsuperscript{th} MFCCs were chosen, which covered the major frequency range of human speech. Thus, each segment was represented by a $40 \times n$ matrix, where $40$ indicated the number of selected MFCCs (the 5\textsuperscript{th} to 44\textsuperscript{th} of 128 MFCCs), and $n$ denoted the number of time bins sized around 10.67msec. The calculation of MFCCs was implemented with LibROSA library \citep{mcfee2015librosa} in Python.

\subsection{Classical Machine Learning Methods}

Classical machine learning methods classify each segment via its features. The feature extraction and selection are essential, determining the performance of the classifier. We examined several commonly used classical machine learning methods in this study, including kNN, RF, and SVM with three different kernels: polynomial kernel, linear kernel, and Gaussian radial basis function (RBF) kernel. These methods were implemented with the Scikit-learn library \citep{scikit-learn} in Python.

\subsubsection{Feature Extraction}

The size of MFCCs matrices varied due to the differences in duration of speech segments. To build fixed-length feature vectors for classical machine learning methods, we compressed the $40 \times n$ matrices along the second dimension (temporal dimension). We calculated four time-domain statistics (average value, standard deviation, skewness, and kurtosis) for each MFCC to encompass the segment-level information, as what \citet{orozco2016towards} had done. To reduce the loss of time-variant information in matrix compression, we also added the first and the second derivatives of MFCCs (MFCCs\textsuperscript{(1)} and MFCCs\textsuperscript{(2)}) and the same time-domain statistics of them. Concatenating all the MFCCs and their derivatives and statistics, we finally aquired a 1-D feature vector of length 480 ($40\ MFCCs \times 3\ derivatives \times 4\ statistics $) for each segment. 

\subsubsection{Feature Selection and Classification}
\label{sec:fs}

We performed feature selection in the entire feature space to remove irrelevant and redundant features. Firstly, we applied nine filtering methods according to \citet{li2017feature} to all features. Among those filtering methods, fisher score, reliefF, and trace ratio were based on similarity. These methods assessed features' importance by their ability to preserve data similarity, especially referring to the data manifold structure encoded by an affinity matrix. RFS, ls\_l21 \citep{liu2012multi}, and ll\_l21 were based on sparse learning. These methods considered minimizing both biases of fitting models and sparse regularization terms. Gini index, f-score, and t-score were based on statistics, assessing feature importance with statistical measures. 
Each method provided a feature-rank, which described the importance of features in classifying patients with PD from HCs. In this sense, features with higher ranks played more significant roles in classification. 

Secondly, we performed feature selection and PD classification with RBF-kernel based SVM classifiers simultaneously. Each time we selected the top $m$ ($m=1,2,...,480$) features according to their ranking orders generated from one of the nine filtering methods as inputs for the classifier. The leave-one-subject-out (LOSO) strategy was deployed during classification: segments from one subject were excluded as test samples, and the rest were used for training. This strategy guaranteed that segments from the same subject only appeared either in the training or test set, eliminating the risk of identity confounding \citep{neto2019detecting}. As a result, LOSO provided a more objective and fair evaluation of classifiers.

We traversed the nine feature-ranks and all possible values of $m$. The classification performance for each of these combinations guided the selection of the best feature subset. We also examined the performance of KNN, RF, and linear- and polynomial-kernel based SVM classifiers, using grid search strategy for parameter tuning.

\subsection{Deep Learning Methods}

Deep learning methods allowed us to use the original MFCCs matrices (see \sectionref{sec:prepro}) as inputs to differentiate PD patients without manually extracting and selecting features, thus avoiding the loss of time-variant information when calculating the four statistics in previous classical machine learning methods. We extracted samples from a sliding window of 40 time bins (which corresponded to the speech signal of about 426.7msec) with an overlap of 75\% for each of the original $40 \times n$ MFCCs matrices. In total, 66438 samples sized $40 \times 40$ were obtained as the inputs for the two types of deep neural networks we designed to perform the classification.

\subsubsection{6-layer Convolutional Neural Network}

CNN has been widely applied in tasks regarding images in various domains. CNN introduces convolutional and pooling layers as hidden layers. 
With its shared-weights architecture, CNN has a great response to the sliding and deforming of images. Accordingly, CNN is suitable for processing the time-frequency representation of speech signals. The CNN architecture proposed here included six layers, described in \tableref{tab:CNN}.

\begin{table}[h]
  \footnotesize
  \caption{Description of the CNN structure}
  \label{tab:CNN}
  \centering
  \begin{tabular}{lll}
    \toprule
    Layer   & Kernel Size           & Output Size\\
    \midrule
    Input   &      \ -                 & $40\times40\times1$\\
    Conv    & $3\times3\times16$    & $40\times40\times16$\\
    Conv    & $3\times3\times16$    & $40\times40\times16$\\
    MaxPool & $2\times2$            & $20\times20\times16$\\
    Conv    & $3\times3\times32$    & $20\times20\times32$\\
    MaxPool & $2\times2$            & $10\times10\times32$\\
    Conv    & $3\times3\times64$    & $10\times10\times64$\\
    MaxPool & $2\times2$            & $5\times5\times64$\\
    MaxPool & $5\times5$            & $1\times1\times64$\\
    Flatten &     \ -                  & 64\\
    FC  & $64 \times 8$           & 8\\
    Dropout & p = 0.5               & - \\
    FC  & $8 \times 2$            & 2\\
    \bottomrule
  \end{tabular}
\end{table}

\subsubsection{Self-attention based Long Short-Term Memory Network}
\label{sec:LSTM}

Long Short-Term Memory (LSTM) networks are a kind of Recurrent Neural Networks (RNNs) dealing with long-term sequences \citep{hochreiter1997long}. Inspired by the physiological process of human decision-making after listening to a speech segment where attention is involved, we constructed a self-attention based LSTM architecture.
The MFCCs matrix was fed into the LSTM layer chronologically frame by frame, producing a 2-D dimensional output matrix V of size $40 \times Hidden\_size$. We designed the attention layer according to the structure of the Transformer proposed by Google \citep{vaswani2017attention}. 
We chose the last output vector as the feature vector extracted by the encoder, fed it into the classifier to make a decision.
The LSTM structure is described in \tableref{tab:RNN}.

\begin{table}[h]
  \footnotesize
  \caption{Description of the LSTM structure}
  \label{tab:RNN}
  \centering
  \begin{tabular}{lll}
    \toprule
    Layer   & Kernel Size       & Output Size\\
    \midrule
    Input   &  -              & $40\times40$\\
    LSTM    & Hidden Size = 512,  &    \\
            & Layer Number = 3   & $40\times512$\\
    Attention & Self-Attention  & $40\times512$\\
    Last output   &    -   & 512\\
    MaxPool & $2$               & 256\\
    FC  & $256\times64$    & 64\\
    FC  & $64\times8$     & 8\\
    Dropout & p = 0.4           & -  \\
    FC  & $8\times2$    & 2\\
    \bottomrule
  \end{tabular}
\end{table}

\subsubsection{End-to-End System}
The end-to-end (E2E) system usually means the network that can utilize the original signal without additional processing. In this work, we also constructed a deep neural network that directly adopted the original waveform signal as the input. We hoped this design could capture preciser time information than using MFCCs as input. We used a time-convolutional layer structure as the front-end network instead of the calculation of MFCCs in \sectionref{sec:prepro}, inspired by the CLDNN structure proposed by \citet{sainath2015learning}. We used the LSTM network as the back-end network to make a final decision. The E2E architecture proposed here is described in \tableref{tab:E2E}.

\begin{table}[h]
  \footnotesize
  \caption{Description of the E2E structure}
  \label{tab:E2E}
  \centering
  \begin{tabular}{lll}
    \toprule
    Layer   & Kernel Size           & Output Size\\
    \midrule
    Input   &       -                 & $1\times 21500$\\
    Unfold &  Window = 21500,  & \\
     &  Stride = 500   &  $40 \times 2000$ \\
    1d-Conv &  $200 \times 40$,  & \\
    & Stride = 25 & $40 \times 73 \times 40$\\
    MaxPool &  $73 \times 1$           & $40 \times 1 \times 40$\\
    Flatten   &       -                 & $40\times 40$\\
    Log-ReLU   &       -                 & $40\times 40$\\
    \midrule
    LSTM    & Hidden Size = 512,  &    \\
            & Layer Number = 3   & $40\times512$\\
    Last output   &    -   & 512\\
    MaxPool & $2$               & 256\\
    FC  & $256\times64$    & 64\\
    Dropout & p = 0.4           & -  \\
    FC  & $64\times8$     & 8\\
    FC  & $8\times2$    & 2\\
    \bottomrule
  \end{tabular}
\end{table}

\subsubsection{Training}

We utilized the same LOSO strategy to divide training and test sets, and further separated 20\% samples from the training set for validation. We chose BCE loss as the loss function, Adam as the optimizer, and set the learning rate to $0.001$. The training process was stopped when accuracy on the validation set reached 0.99 or the maximum number of epochs reached (in this case, we chose the number of epochs when the highest validation accuracy was achieved).

\subsection{Evaluation}

We evaluated the performances of classifiers at segment-level. The classical machine learning methods directly provided a segment-level classification. While in deep learning methods, we fed the $40 \times 40$ MFCCs matrix (named 'sample' of a speech 'segment') into the classifier, thus made a sample-level classification. We then averaged the prediction probabilities of samples belonged to the same segment. The segment's label was assigned according to the averaged probability. To improve robustness and generalizability,  we excluded samples with the top and bottom 30\% probability values from averaging. This strategy generated a segment-level classification result, making it available to compare the performance of deep learning and classical machine learning methods.

The classifiers were evaluated through the prediction accuracy (ACC) by calculating the ratio of correctly classified segments to all segments. Confusion matrix and Area Under the receiver operating characteristics Curve (AUC) were also used as the extension evaluation criterion.\

\begin{figure}[h]
\centering
\includegraphics[scale=0.5]{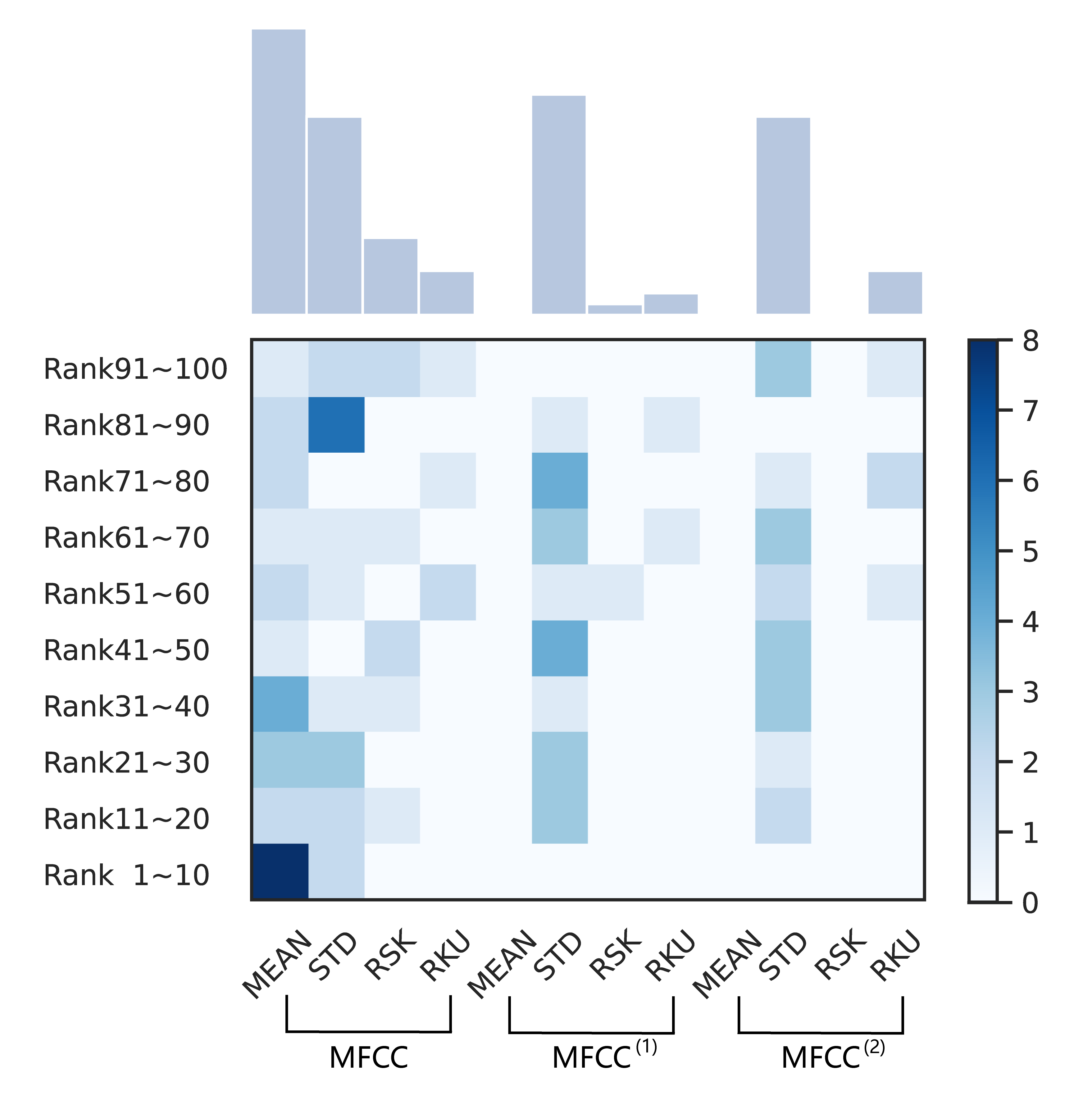}
\caption{Distribution of selected features}
\label{fig:dis}
\end{figure}

\section{Results and Discussion}
\label{sec:vec}

\begin{figure*}[ht]
\centering
\includegraphics[scale=0.4]{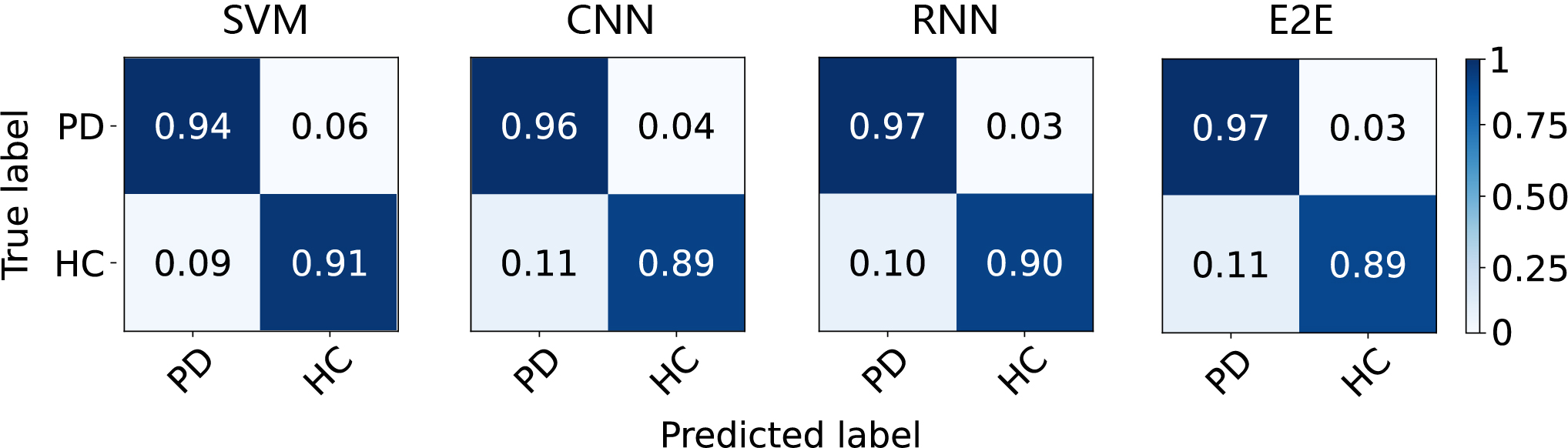}
\caption{Confusion matrices of four methods}
\label{fig:cm}
\end{figure*}

For classical machine learning methods, using ls\_121, the sparse learning-based feature selection method led to the achievement of the best performance comparing to others. We selected the top 100 features with ls\_l21 as the input feature subset for classification. This group of features reached execellent performance (exceeded 99\% of maximum AUC) with a small number of features. \figureref {fig:dis} presents the distribution of this feature subset.
MFCCs accounted for a large proportion of selected features at higher ranks, whereas the proportion of MFCCs\textsuperscript{(1)} and MFCCs\textsuperscript{(2)} raised in lower rank intervals. The total number of each statistic selected for classification is shown in the upper part of \figureref{fig:dis}. The average value of MFCCs played a major role in classification, indicating the lower speech volume of PD patients. The widely distributed standard deviation of MFCCs and its derivatives indicated the importance of time-variant information in speech signals. This result revealed the contribution of each feature in differentiating the PD patients.

Among the three classical machine learning classifiers mentioned in \sectionref{sec:fs}, RBF-kernel based SVM with the selected feature subset achieved the highest AUC of 0.978 and ACC of 0.930, which was significantly better than those acquired with all 480 features (AUC = 0.966 and ACC = 0.906), showing the importance of feature selection. \figureref{fig:cm} (a) shows the confusion matrix for RBF-kernel based SVM, with both the sensitivity and specificity higher than 0.89. After grid search, the best parameters we found in RBF-kernel were C = 4 and gamma = $1/(\# features \times \ variance \ of \ training \ samples)$.

Other classifiers with the top 100 feature subset provided by ls\_l21 also performed well, though worse than RBF-kernel SVM: AUC = 0.955 and ACC = 0.880 for RF, AUC = 0.923 and ACC = 0.868 for kNN, AUC = 0.931 and ACC = 0.854 for linear kernel based SVM, and AUC = 0.954 and ACC = 0.875 for polynomial kernel based SVM.

\begin{figure}[h]
\centering
\input{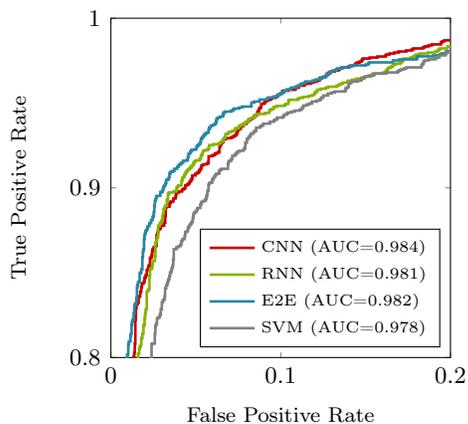}
\caption{ROC curve of four methods}
\label{fig:ROC}
\end{figure}

For deep learning methods, the 6-layer CNN, self-attention based LSTM and E2E system all exhibited better performance. CNN performed the best, with an AUC of 0.984 and ACC of 0.938. Self-attention based LSTM reached an AUC of 0.981 and ACC of 0.942, while E2E system reached an AUC of 0.982 and ACC of 0.945. \figureref{fig:cm} (b), (c) and (d) presents the confusion matrices for these three methods respectively. The better result may due to the detailed usage of signal information. In classical machine learning methods, we roughly represented the signal by the statistics of MFCCs, losing some time-variant information of the speech signal. \citet{orozco2015voiced} successfully classify PD speech from voiced/unvoiced transitions, showing that parkinsonian speech disorder can be detected from the rapid changes of speech. The original MFCCs matrix can provide a higher time-domain resolution than its statistics, which may benefit the classification. The local view of each method's ROC curve is presented in \figureref{fig:ROC}, providing a visual comparison of these four methods.

\begin{figure}[hb]
\centering
\input{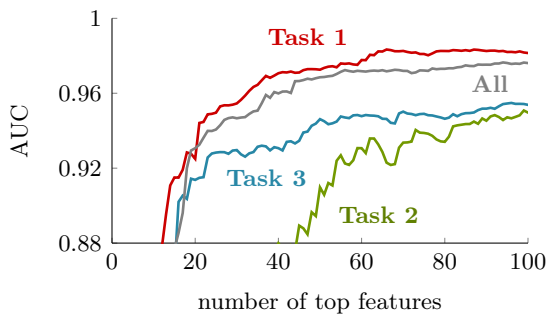}
\caption{AUC of each tasks}
\label{fig:AUC_tasks}
\end{figure}

This study included three speech tasks, as described in \sectionref{sec:data}. Accordingly, we tested the classification performance on each task individually. The same feature selection procedure in \sectionref{sec:fs} was applied. \figureref{fig:AUC_tasks} shows the AUC calculated for classification performance with different numbers of selected features using RBF-kernel based SVM.

The AUC of task 2 (DDK) is lower than task 1 (image description) and task 3 (text reading), which is consistent with the result shown in \citet{orozco2016automatic}. Sentences involved different words, pauses, and rhythms, while DDK in this study mainly contained the single vowel /a/. Thus DDK covered less variation in voice and only reflected a narrower range of vocal space. This performance difference might support the assumption that MFCCs are more suitable for complicated tasks or speech, as discussed in \citet{moro2020using}.

Besides, task 1 achieved a significantly better classification result compared to task 3. The reason might be that text reading only involved fixed contents, while image description allowed more freedom with fewer limitations. Another possible explanation of this difference could be that subjects were more focused on recalling the image's content when they participated in task 1, thus relaxed the control of some muscles and exhibited more severe speech disorder. As a result, the difference between PD patients and healthy subjects' speech was shown more obviously. This trick of attracting patients' attention is also used in other tests to illuminate their movement disorder. PD patients could be classified from natural speech without specifically designed tasks, indicating the potential to detect PD patients from speech obtained in daily life and assist early clinical diagnosis of PD.

Though worse than task 1, task 3 also achieved a strong classifying ability and was better than task 2. The reading texts we designed in task 3 were two ancient Chinese poems, which are very familiar to Chinese people. Subjects could read the poems in their most comfortable way as a conditional reflex, significantly reduced the time cost. Besides, it is uncertain whether image description may be influenced by cognitive function. Reading tasks were more focused on speech function and can provide an equal number of training data, which is suitable for machine learning.

\begin{figure*}[h]
\centering
\includegraphics[scale=0.5]{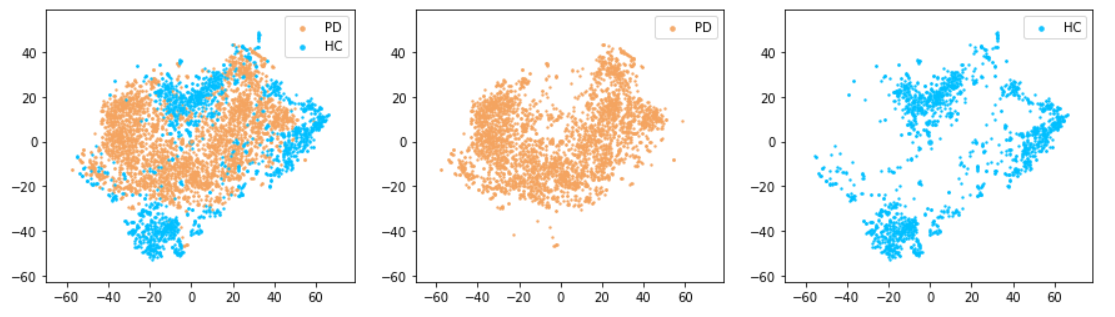}
\caption{t-SNE result of selected features}
\label{fig:tsne}
\end{figure*}

We visualized the selected features using t-distributed stochastic neighbor embedding (t-SNE). In detail, we performed principal component analysis (PCA) to reduce the features from 480-d to 50-d and retain more than 90\% of the effective information. Then we performed t-SNE on PCA result. \figureref{fig:tsne} showed the separability of features from the two groups.

We further compared our RBF kernel-based SVM results in single speech task with previous studies, as shown in \tableref{tab:compare}. Our method (in bold) surpassed previous state-of-the-art results.

\begin{table*}[hb]
  \footnotesize
  \caption{ACC comparison with previous studies}
  \label{tab:compare}
  \centering
  \begin{threeparttable}
  \begin{tabular}{@{}llccc}
    \toprule
    \multirow{2}{*}{\textbf{Article\ \ \ \ \ \ \ \ \ \ \ \ \ \ \ \ \ \ \ \ \ \ }} & \multirow{2}{*}{\textbf{Language\ \ \ }} &    \multicolumn{3}{c}{\textbf{Vocal task}} \\
    & & \textbf{Talking}\tnote{1}  & \textbf{\ \ DDK\ \ }  & \textbf{Reading}
    \\
    \midrule
    \citet{hauptman2019identifying} & Hebrew & 0.741 & 0.741 & 0.721\\
    \citet{jeancolas2019comparison} & French & 0.74 & 0.78 & -\\
    \citet{lopez2019assessing} & Spanish & 0.84 & 0.78 & 0.80\\
    \citet{vaiciukynas2017parkinson} & Lithuanian & - & - & 0.859\\
    Ours  & Chinese  & \textbf{0.940} & \textbf{0.835} &    \textbf{0.911}\\
    \bottomrule
  \end{tabular}
  \begin{tablenotes}
    \footnotesize
    \item[1] Talking task includes free talk, monologue, and image description
  \end{tablenotes}
  \end{threeparttable}
\end{table*}

\section{Conclusions and Future Work}
\label{sec:floats}

In this study, we built a speech dataset with different tasks in Mandarin Chinese; investigated features for speech disorders related to Parkinson's disease; developed high-accuracy methods for the classification of patients with PD and healthy subjects.

Both classical machine learning methods with feature selection and deep learning methods we developed have state-of-the-art performance. In addition, we tested the classification performance for different speech tasks. The result suggests that free talk has stronger classification power than standard tasks, which could aid the design of future speech tests for efficient early diagnosis of the disease.

For future work, more specifically designed neural networks may further improve accuracy and generalizability. The pipeline of feature selection and the neural network structure proposed in this study could also be applied to speech tasks in other languages. High quality and large volume data on speech tasks from patients and healthy subjects are crucial for practical applications. A better understanding of the speech disorders' mechanism would also help us in developing more effective diagnostic tools. 

\acks{This work is sponsored by The National Key Research and Development Program of China (2016YFC0105502), NSFC (81527901), and Shenzhen International Cooperative Research Project (GJHZ20180930110402104). All the authors are affiliated with the National Engineering Laboratory for Neuromodulation, School of Aerospace Engineering, Tsinghua University. Luming Li is also affiliated with: Precision Medicine \& Healthcare Research Center, Tsinghua-Berkeley Shenzhen Institute, Tsinghua University; IDG / McGovern Institute for Brain Research, Tsinghua University; Institute of Epilepsy, Beijing Institute for Brain Disorders. Correspondence to Yanan Sui and Luming Li.}


\bibliography{jmlr-sample}


\end{document}